\documentclass{kapproc}
\usepackage[dvips]{graphicx}
\upperandlowercase
\kluwerbib 
\let\lcitebracket(
\let\rcitebracket)
\begin{document}
%
%
%
\articletitle[Three-Dimensional Bar Structure]{Three-Dimensional Bar Structure and Disc/Bulge Secular Evolution}
\chaptitlerunninghead{Three-Dimensional Bar Structure}
\author{M.\ Bureau\altaffilmark{1}, G.\ Aronica\altaffilmark{2,3}, E.\ Athanassoula\altaffilmark{3}}

\altaffiltext{1}{Sub-Department of Astrophysics, University of Oxford,
  Denys Wilkinson Building,\\Keble Road, Oxford OX1~3RH, United Kingdom}
\altaffiltext{2}{Astronomisches Institut, Ruhr-Universit\"{a}t Bochum,
 D-44780 Bochum, Germany}
\altaffiltext{3}{Observatoire de Marseille, 2 place Le Verrier,
  F-13248 Marseille Cedex~4, France}
%
%
\begin{abstract}
$K\!$n-band imaging of a sample of $30$ edge-on spiral galaxies with a
boxy or peanut-shaped (B/PS) bulge is discussed. Galaxies with a B/PS
bulge tend to have a more complex morphology than galaxies with other
bulge types, unsharp-masked images revealing structures that trace the
major orbit families of three-dimensional bars. Their surface
brightness profiles are also more complex, typically containing $3$ or
more clearly separated regions, including a shallow or flat
intermediate region (Freeman Type~II profiles), suggestive of
bar-driven transfer of angular momentum and radial redistribution of
material. The data also suggest abrupt variations of the discs'
scaleheights, as expected from the vertical resonances and
instabilities present in barred discs but contrary to conventional
wisdom. Counter to the standard `bulge + disc' model, we thus propose
that galaxies with a B/PS bulge are composed of a thin concentrated
disc (a disc-like bulge) contained within a partially thick bar (the
B/PS bulge), itself contained within a thin outer disc. The inner disc
most likely formed through bar-driven processes while the thick bar
arises from buckling instabilities. Both are strongly coupled
dynamically and are formed mostly of the same (disc) material.
\end{abstract}
%
%
\section{Introduction}
Bulges are traditionally viewed as low-luminosity elliptical galaxies,
suggesting a rapid formation dominated either by mergers/accretion
(e.g.\ Searle \& Zinn 1978) or possibly by dissipative gravitational
collapse (e.g.\ Eggen et al.\ 1962). Those ideas have come under
increasing criticism, however, and competing models where bulges grow
secularly (i.e.\ over a long timescale and in relative isolation) are
now widely discussed, many of them bar-driven (e.g.\ Pfenniger \&
Norman 1990; Friedli \& Benz 1995).

We focus here on the identification of most boxy and peanut-shaped
(B/PS) bulges in edge-on spiral galaxies with part of the bars of
barred spirals. $N$-body simulations clearly show that, whenever a
disc forms a bar, a B/PS bar/bulge develops soon after (e.g.\ Combes
\& Sanders 1981; Combes et al.\ 1990). True peanuts are seen with the
bar nearly side-on while boxier/rounder shapes are seen when the bar
is closer to end-on. This view is supported by the incidence of B/PS
bulges in edge-on spirals (e.g.\ L\"{u}tticke et al.\ 2000a) and by
the ionized-gas and stellar kinematics of discs harbouring a B/PS
bulge (e.g.\ Kuijken \& Merrifield 1995; Bureau \& Freeman 1999; Chung
\& Bureau 2004).

Following recent work by L\"{u}tticke et al.\ (2000b), we present here
additional evidence for the above scenario based on $K\!$n-band
imaging of a sample of well-studied nearby edge-on spiral galaxies
with a B/PS bulge. The observations and results are described more
deeply in Bureau et al.\ (2006) and Athanassoula et al.\ (2006).
%
%
\section{Images}
\begin{figure}[t]
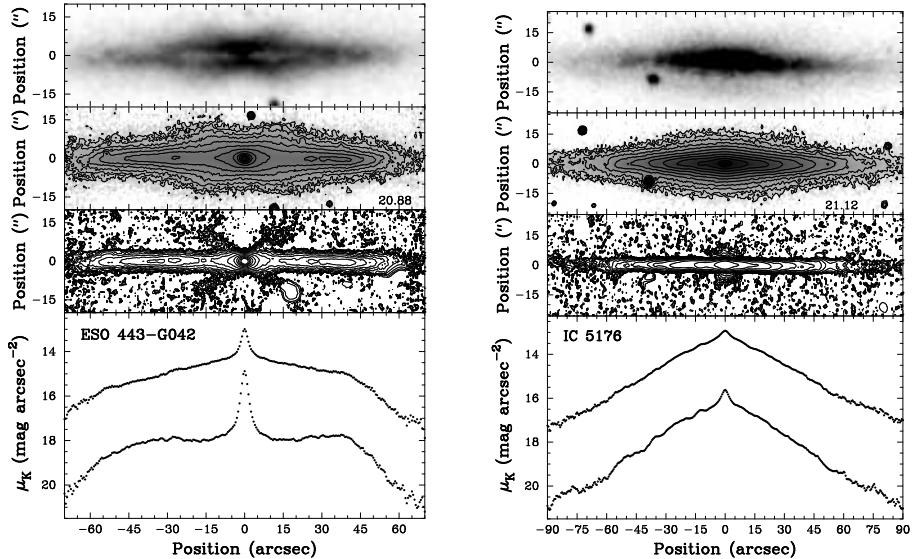

\begin{center}
\includegraphics[width=0.465\textwidth]{fig1a.ps}
\hspace*{0.05\textwidth}
\includegraphics[width=0.465\textwidth]{fig1b.ps}
\end{center}
\vspace*{-3mm}
\caption{Images and surface brightness profiles of the galaxies
 ESO443-G042 (left), with a B/PS bulge, and IC5176 (right), with a
 nearly pure disc. From top to bottom, each panel shows first a DSS
 image of the galaxy, second our $K\!$n-band image, third an
 unsharp-masked $K\!$n-band image, and last the major-axis (fainter)
 and vertically-summed (brighter) surface brightness profiles, all
 spatially registered.}
\label{fig:images}
\end{figure}

The $30$ edge-on spiral galaxies of Bureau \& Freeman (1999) and Chung
\& Bureau (2004), $24$ of which have a B/PS bulge (the rest
constituting a control sample), were observed at $K\!$n-band using
{\small CASPIR} on the 2.3m telescope at Siding Spring
Observatory. Both a standard image and numerous unsharp-masked images
(enhancing local extrema) were produced for every galaxy. Examples are
shown in Figure~\ref{fig:images} for a B/PS bulge and a nearly
bulgeless galaxy. Compared to other bands, $K\!$n is least hampered by
dust and best traces the dominant Population~II stars, sharpening the
B/PS and associated features.

The unsharp-masked images highlight pervasive complex morphological
structures, such as centered X features, off-centered X features,
secondary maxima along the major-axis, elongated minor-axis extrema,
spiral arms, etc (see Fig.~\ref{fig:images}). Most importantly, except
for the minor-axis extrema, those structures are much more prevalent
in galaxies with a B/PS bulge. For examples, $88\%$ of galaxies with a
B/PS bulge have either a centered or off-center X feature, while only
$33\%$ of the control galaxies do, with identical fractions for
secondary major-axis maxima. The contrast between the main and control
samples would in fact be even greater if the latter was not
contaminated by weak B/PS bulges. Although the accretion of external
material can give rise to centered X-shaped features (e.g.\ Hernquist
\& Quinn 1988), it is unlikely that long-lasting off-centered X could
be produced, and those are the majority of the features observed in
our sample. The orbital structure of 3D bars offers a more attractive,
simple and unifying and explanation.
%
%

The most important orbit families in 3D bar models are those of the
$x_1$ tree, elongated parallel to the bar and located within
corotation. This includes the $x_1$ family itself (restricted to the
equatorial plane) and other families bifurcating from it at vertical
resonances. The morphological features observed can be reproduced by
superposing orbits of the appropriate shapes, as done by Patsis et
al.\ (2002). This is particularly true of the (off-)centered X
features which, depending on the model (mass distribution and pattern
speed) and viewing angle, can {\em both} arise from orbit families
extending vertically out of the equatorial plane. The same orbits can
give rise to a number of maxima along the major-axis, similar to the
secondary maxima observed. Those generally occur at larger radii than
the X features and near (but within) the ends of the bar. An analogous
explanation is that the secondary maxima are the edge-on projections
of the inner rings present in a large fraction of barred spiral
galaxies (e.g.\ Buta 1995) and predicted to form under the influence
of bars in gas-rich and gas-poor discs (e.g.\ Schwartz 1981;
Athanassoula \& Misiriotis 2002). Unsharp-masking of barred $N$-body
simulations also provides a perfect match to the variety of features
observed, strengthening the link between them and edge-on bars
(Athanassoula 2005).
%
%
\section{Surface Brightness Profiles}
We also extracted from our images standard major-axis surface
brightness profiles and profiles summed in the vertical direction (as
if the galaxies were infinitely thin). From axisymmetric face-on
galaxies, we would generally expect the profiles to show only two
distinct regions: a first steep region at small radii, associated with
the bulge, and a second exponential region at larger radii, associated
with the disc. Such profiles are however rare in our data, especially
along the major-axis.

In particular, the profiles of galaxies with a B/PS bulge are again
more complex than those of the control sample, in that they typically
contain more distinct regions separated by clear radial breaks. For
example, $96\%$ of the galaxies with a B/PS bulge have a major-axis
profile with an additional region at intermediate radii, where the
profile is very shallow, even flat or slightly rising (Freeman Type~II
profile; see Fig.~\ref{fig:images}). The fraction for the control
sample is only $50\%$, and the contrast between the two samples would
again be sharper if they had been better selected.

Those shallow intermediate regions are particularly important as they
suggest a third photometric/morphological component at moderate radii,
inconsistent with a classic axisymmetric bulge + disc model. Both the
central peak and the flat intermediate region are however consistent
with a {\em single} bar viewed edge-on, with no need for a classic
bulge. Indeed, Bureau \& Athanassoula (2005) followed the evolution of
the major-axis profile in barred $N$-body simulations viewed edge-on,
and they convincingly showed that bar formation and evolution is
associated with the buildup and continued growth of a dense central
region, which would normally be identified with a bulge, and with the
formation and gradual flattening of an intermediate region, in
addition to the outer exponential disc. The intermediate region
extends to the end of the bar, well beyond the central peak and the
thickest part of the bar, as observed.

As expected from the elongated boxy/peanut shape of the bar in
simulations, the observed ratio of the length of the thickest part of
the bulge (or the central peak) to that of the flat intermediate
region is also generally larger in peanut-shaped bulges than in boxy
ones. There is however much variety, and likely many causes for
it. Even so, the scatter in the ratio may well be dominated by the
range of bar strengths in the sample, rather than by the range of
viewing angles, as also suggested by the data of L\"{u}tticke et al.\
(2000b). The small ratios observed in strong peanut-shaped bulges (see
Fig.~\ref{fig:images}) can then be explained only if the central peak
and the thick part of the bulge are shorter in stronger bars. This is
natural in barred models if the central peak is a disc-like bulge
limited by the outer inner Lindblad resonance (e.g.\ Athanassoula
1992).
%
%
\section{Bar-Driven Evolution}
Athanassoula (2003) showed that much of the bar-driven evolution in
discs is due to a transfer of angular momentum from the inner (barred)
disc to the outer disc and halo, leading to a {\em radial}
redistribution of matter. The majority of vertically-summed surface
brightness profiles, best suited to isolate those effects, indeed show
$3$ or more clearly separated regions, while only one of the control
galaxy does. Collapse or accretion/merger scenarios can not
straightforwardly create those radial breaks or explain their spatial
correlation with the ionized-gas and stellar kinematics. But if
bar-driven scenarios are right, the break at the end of the
intermediate region should mark the bar's end. Comparison shows that
it indeed occurs where the rotation curve flattens, normally
associated with the end of the bar. The break also coincides with the
inner ring, when visible. As inner rings occur near the inner 4:1 and
corotation resonances (e.g.\ Schwartz 1981), our galaxies are
consistent with harbouring fast bars, as do most galaxies (e.g.\
Gerssen et al.\ 2003).

To probe the {\em vertical} redistribution of material predicted by
bar buckling scenarios, we must compare the major-axis and
vertically-summed profiles. If the stellar scaleheight was constant
with radius, the two profiles would have the same functional form but
different zero-points (i.e.\ be offset but parallel). This is clearly
not the case for most galaxies, however, and the profiles of most
galaxies with a B/PS bulge differ significantly (e.g.\
Fig.~\ref{fig:images}). Although we have amalgamated the bulge and
disc by considering a single scaleheight, the functional difference
between the two profiles is greatest in the flat intermediate regions,
which are clearly disc material dominated. Our profiles thus clearly
show that the radial scaleheight variations are real and that they
occur in the discs, in direct contradiction to the common wisdom that
disc scaleheights are constant (e.g.\ van der Kruit \& Searle
1981). Athanassoula et al.\ (2005) show that the variations are as
expected from barred $N$-body models.

Galaxy bulges are usually defined as either 1) the steep central
component of the surface brightness profile or 2) the thick galactic
component. Those definitions are normally used interchangeably, but
many results show this to be grossly oversimplified (e.g.\ Kormendy \&
Kennicutt 2004). Our data clearly show that the central peak is often
contained {\em within} the thick central component, while the shallow
intermediate region always extends {\em beyond} the thick component
(e.g.\ Fig.~\ref{fig:images}). Those facts are unaccounted for in
classic bulge formation scenarios, but they are a natural consequence
of the bar viewing angle and the fact that only part of the bar is
actually thick in bar-buckling models (see Athanassoula 2005 for more
on this last point).

Comparison of the major-axis and vertically-summed profiles also
reveals that the central peak is more pronounced along the major-axis,
so that most of the high $z$ material belongs to the shallow
intermediate region rather than the central peak. The latter thus
seems to be a thin concentrated disc, while the former appears to be
thick. This is again counter to the classic bulge + disc model, but
fits with the nomenclature proposed by Athanassoula (2005). The bar
leads to the formation of a concentrated disc (a disc-like bulge),
presumably through (bar-driven) gaseous inflow and star formation, but
this disc is thin, largely decoupled from the bar, and addresses only
the first bulge definition. The bar itself is thick over most but not
all its length (a B/PS bulge), with a shallow profile, and addresses
the second bulge definition. Like the classic models, our model
comprises a number of distinct building blocks, but those are very
different and tightly intertwined dynamically, emerging from the rapid
radial variation of the scaleheight of the disc material, due to the
weak but relentless action of bar-related resonances.
%
%

%
\end{document}